\documentclass[12pt]{iopart}
\usepackage[usenames,dvipsnames]{color}
\usepackage{graphicx}
\usepackage{iopams}

\begin{document}
\title{Quantum analogue of energy equipartition theorem}
\author{P. Bialas, J. Spiechowicz and J. {\L}uczka}
\address{Institute of Physics and Silesian Center for Education and Interdisciplinary Research, University of Silesia, 41-500 Chorz{\'o}w, Poland}
\ead{j.spiechowicz@gmail.com} 
\begin{abstract} 
One of the fundamental laws of classical statistical physics is the energy equipartition  theorem which states that for each degree of freedom the mean  kinetic energy $E_k$ equals $E_k=k_B T/2$, where $k_B$ is the Boltzmann constant and $T$ is temperature of the system. Despite the fact that quantum mechanics has already been developed for more than 100 years still there is no quantum counterpart of this theorem. We attempt to fill this far-reaching gap and consider the simplest system, i.e.  the Caldeira-Leggett model for a free quantum Brownian particle in contact with thermostat consisting of an infinite number of harmonic oscillators. We prove  that the mean kinetic energy $E_k$ of the Brownian particle equals the  mean  kinetic  energy $\langle \mathcal E_k \rangle$ per one degree of freedom  of the thermostat oscillators, i.e. $E_k = \langle \mathcal E_k \rangle$.  
We show that this relation can be obtained from the fluctuation-dissipation theorem derived within the linear response theory and is universal in the sense that it  holds true for any linear and non-linear systems in contact with bosonic thermostat. 
\end{abstract}
\maketitle
The energy equipartition theorem (EET) is one of the turning points in the history of the development of classical statistical physics. Already in 1845 John James Waterston, an often forgotten pioneer of the kinetic theory of gases, proposed equipartition of kinetic energy for translational motion \cite{waterston}. 
This idea was further extended by the fathers of modern statistical physics in the persons of James Clerk Maxwell and Ludwig Boltzmann. The latter in 1876 showed that average kinetic energy is equally shared in a portion of $E_k = k_B T/2$ among all degrees of freedom of a system \cite{boltzmann}. Then he provided the first and one of the most remarkable application of this theory in the form of explanation of Dulong-Petit law for the specific heat capacities of solids. Since that time the EET has become one of the most important and most useful relation exploited in various branches of Natural Science, including physics, chemistry and even biology. 
For classical systems EET represents a universal relation in the sense that it does not depend on a number of particles in the system, a potential force which acts on them, any interaction between particles or the strength of coupling between the system and thermostat \cite{huang}. In contrast, this is no longer true for quantum systems. To illustrate this problem we present the clich{\'e}d example taken form the classic book by Feynman \cite{feynman}, i.e. a harmonic oscillator of mass $m$ and eigenfrequency $\omega_0$ for which the Hamiltonian has the well-known form $H = p^2/2m + m \omega_0^2 x^2/2$. In this case the average kinetic energy $\mathcal{E}_k$ reads \cite{feynman}
\begin{equation}\label{ho}
\mathcal{E}_k(\omega_0) = \frac{1}{2m} \langle p^2 \rangle = 
\frac{\hbar \omega_0}{4} \coth{\frac{\hbar \omega_0}{ 2k_BT}}
\end{equation} 
and evidently depends on the potential form via the frequency $\omega_0$. The limit $\omega_0 \to 0$ corresponds to the Hamiltonian of a free particle for which one gets the classical results  $\mathcal{E}_k (\omega_0=0) = k_B T/2$. Eq. (\ref{ho}) is  derived for a system in a Gibbs canonical state and in  consequence is valid only in the weak coupling limit between the oscillator and the thermostat. If it does not hold the problem is much more complicated. Moreover, in many cases the EET has been exploited far away from its domain of applicability and range of its validity, e.g. from this relation it follows that at zero temperature $T=0$ the kinetic energy is zero. It is obviously not true because even at $T=0$ there are quantum vacuum fluctuations of the thermostat.

From the time of Max Planck and birth of quantum physics  a 
quantum counterpart  of EET has not been explicitly formulated. In literature, both in original papers and text-books, there are various  expressions for  mean energy of some particular systems, see e.g. Refs.  \cite{hakim,ford,ingold,jasnow,lampo2,breuer,weis}.  
Here, we keep things maximally simple to present our genuine result in a transparent way. Therefore we study the most basic model of a quantum open system in the form of a one-dimensional quantum Brownian motion. Hereafter, we consider a particle of mass $M$ subjected to the potential $U(x)$ and interacting with a large number of independent oscillators which form the thermostat (environment) of temperature $T$ being in an equilibrium Gibbs canonical state. The  Hamiltonian of such a total system can be written as \cite{maga,ingold,breuer,weis,hujcen,chaos,ph} 
(a complete list of papers concerning this problem is too long and inevitably our choice is subjective.)  
\begin{equation} \label{H}
H=\frac{p^2}{2M}+U(x) + \sum_i \left[ \frac{p_i^2}{2m_i} + \frac{m_i\omega_i^2}{2} \left( q_i - \frac{c_i x}{m_i \omega_i^2} \right)^2 \right],
\end{equation}
where the coordinate and momentum operators $\{x, p\}$ refer to the Brownian particle and $\{q_i, p_i\}$ are the coordinate and momentum operators of the $i$-th heat bath oscillator of mass $m_i$ and the eigenfrequency $\omega_i$. The parameter $c_i$ characterizes the interaction strength of the particle with the $i$-th oscillator. 
All coordinate and momentum operators obey canonical equal-time commutation relations. From the Heisenberg equations of motion for all coordinate and momentum operators one can obtain an effective equation of motion for the particle coordinate operator $x(t)$. It is called a generalized Langevin equation (GLE). Again, to make our work maximally transparent we now consider an elementary example, namely, a free Brownian particle for which $U(x) = 0$ and the GLE reads
\begin{equation}\label{GLE}
M{\ddot x}(t)+\int_0^t du \; \gamma(t-u) \dot{x}(u) = -\gamma(t) x(0)+ \eta(t), 
\end{equation}
where $\gamma(t)$ is a dissipation function (damping or memory kernel) and $\eta(t)$ denotes the random force, 
\begin{eqnarray} 
\gamma(t) &=& \int_0^{\infty} d \omega J(\omega) \cos(\omega t),   \label{g}\\
\eta(t) &=& \sum_i c_i \left[q_i(0) \cos(\omega_i t) + \frac{p_i(0)}{m_i \omega_i}
\sin(\omega_i t) \right].
\label{force} 
\end{eqnarray}  
$J(\omega)$ is a spectral density of thermostat modes which contains all information on the system-thermostat coupling:
\begin{equation}\label{spectral}
J(\omega) = \sum_i \frac{c_i^2}{m_i \omega_i^2} \delta(\omega -\omega_i). 
\end{equation}

In the standard approach it is assumed that the initial state $\rho(0)$ of the total system is uncorrelated, i.e., \mbox{$\rho(0)=\rho_S(0)\,\otimes\,\rho_T(0)$}, where $\rho_S$ is an arbitrary state of the Brownian particle and $\rho_T$ is an equilibrium Gibbs canonical state of thermostat of temperature $T$. Next, the thermodynamic limit is imposed meaning that the thermostat is infinitely extended and the singular spectral function $J(\omega)$ in Eq. (\ref{spectral}) approaches a (piecewise) continuous function. The operator-valued random force $\eta(t)$ is a family of non-commuting operators whose commutators are $c$-numbers. Its mean value is zero, $\langle \eta(t) \rangle \equiv \mbox{Tr}\left[\eta(t)\rho_T\right] = 0$ and the symmetrized correlation function $C(t_1-t_2) = (1/2) \langle \eta(t_1)\eta(t_2)+\eta(t_2)\eta(t_1)\rangle$ is related to the memory kernel $\gamma(t)$ via the famous fluctuation-dissipation theorem  \cite{call,kub}
\begin{equation}\label{fludiss}
\hat{C}_F(\omega) = \frac{\hbar \omega}{2} \coth{\left( \frac{\hbar \omega}{2 k_B T} \right)} \hat{\gamma}_F(\omega) = 2\,\mathcal{E}_k(\omega) \hat{\gamma}_F(\omega),
\end{equation}
where $\hat{C}_F(\omega)$ and $\hat{\gamma}_F(\omega)$ are the Fourier cosine transforms of the correlation function $C(t)$ and the dissipation function $\gamma(t)$, respectively 
($\hat{g}_F(\omega) = (2/\pi) \int_0^\infty dt\,g(t) \cos(\omega t)$ for any even function $g(t)$).

To calculate the average kinetic energy \mbox{$E_k= \lim_{t\to\infty} \langle p^2(t) \rangle/2M$} of the Brownian particle we solve the GLE (\ref{GLE}) by the Laplace transform method and find the momentum  
\begin{eqnarray}\label{p(t)} 
p(t) = \mathbf{R}(t)p(0) + \int_0^t du\; \mathbf{R}(t-u) \gamma(u)x(0) + \int_0^t du\; \mathbf{R}(t-u) \eta(u), 
\end{eqnarray}
where the response function $\mathbf{R}(t)$ is determined by its Laplace transform 
\begin{equation}\label{RL} 
\hat{\mathbf{R}}_L(z) = \frac{M}{Mz + \hat \gamma_L(z)}, \quad
\hat \gamma_L(z) = \int_0^{\infty} dt \; {\mbox e}^{-zt} \gamma(t)
\end{equation}  
and $\hat \gamma_L(z)$ is the Laplace transform of the dissipation function $\gamma(t)$. By use of (\ref{p(t)}) we construct the symmetrized momentum-momentum correlation function $(1/2)\langle p(t)p(s) +p(s)p(t)\rangle$, utilize the fluctuation-dissipation relation (\ref{fludiss}), take $t=s$ and perform the limit $t\to\infty$, we finally obtain the expression for the average kinetic energy in the form
\begin{equation}\label{Ek}
E_k = \langle \mathcal{E}_k \rangle = \int_0^{\infty} d\omega \; \mathcal{E}_k(\omega)\mathbb{P}(\omega),  
\end{equation}
where $\mathcal{E}_k(\omega)$ is given by Eq. (\ref{ho}) and
\begin{equation}\label{P}
\mathbb{P}(\omega) = \frac{1}{\pi} \left[\hat{\mathbf{R}}_L(i\omega) + \hat{\mathbf{R}}_L(-i\omega) \right]. 
\end{equation}
The formula (\ref{Ek}) together with Eq. (\ref{P}) constitutes a quantum analogue  for partition of kinetic energy which we formulate in the following way: \emph{The mean kinetic energy $E_k$ of the Brownian particle is equal to mean  kinetic energy per one degree of freedom of the thermostat free oscillators}. The averaging is twofold:  (i) over the Gibbs canonical state $\rho_T$ for the thermostat free (non-interacting with the Brownian particle) oscillators resulting in $\mathcal{E}_k(\omega)$ given by \mbox{Eq. (\ref{ho})} and (ii) over frequencies $\omega$ of  those thermostat oscillators which contribute to $E_k$  according to the probability distribution   $\mathbb P(\omega)$. 
By  Eqs. (\ref{P}), (\ref{RL}) and  (\ref{g}),  the function 
$\mathbb{P}(\omega)$ depends on the spectral density $J(\omega)$ and in consequence  the thermostat oscillators of various frequencies contribute in a greater or lesser degree to the kinetic energy  $E_k$ of the Brownian particle. We discuss this aspect in the latter part of the paper.

\textbf{Theorem}: The function $\mathbb{P}(\omega)$ defined by Eq. (\ref{P}) is a probability measure on a positive half-line of real numbers meaning that
\renewcommand{\theenumi}{\Alph{enumi}}
\begin{enumerate}
	\item $\mathbb{P}(\omega) \ge 0$, 
	\item $\int_0^{\infty} d\omega \; {\mathbb P}(\omega) = 1.$
\end{enumerate}

\textbf{Proof}: (A) First we prove the non-negativity of ${\mathbb P}(\omega)$. Using the definition (\ref{RL}) the function ${\mathbb P}(\omega)$ can be rewritten in the form 
\begin{equation} \label{Pp}
\mathbb{P}(\omega) = \frac{2 M}{\pi} \frac{ A(\omega)}{A^2(\omega)+[B(\omega)-M\omega]^2}.
\end{equation}
We applied the relation $\hat{\gamma}_L(i\omega) = A(\omega) - i B(\omega)$ with
\begin{eqnarray} 
A(\omega) &=&  \int_0^{\infty} dt \; \gamma(t) \cos{(\omega t)},  \label{cos} \\
B(\omega) &=& \int_0^{\infty} dt \; \gamma(t) \sin{(\omega t)}. \label{sin}
\end{eqnarray}
The denominator in (\ref{Pp}) is always positive and it is sufficient to show that the numerator $A(\omega) \ge 0$. 
From Eq. (\ref{g}) we infer that $A(\omega)=  (\pi/2) J(\omega)$. From Eq. (\ref{spectral}) it follows that  $J(\omega) \ge 0$ and the same holds true in the thermodynamic limit when $J(\omega)$ becomes a (piecewise) continuous function. Therefore $\mathbb{P}(\omega) \ge 0$.

(B) Now we prove the normalization condition. 
The function $\mathbb{P}(\omega)$  defined by Eq. (\ref{P}) is an even function and from (\ref{P}) one can obtain its equivalent form 
\begin{equation} \label{PC}
 \mathbb{P}(\omega) = \frac{2}{\pi} \int_0^{\infty} dt\; \mathbf{R}(t) \cos(\omega t) 
 = \hat{\mathbf{R}}_F(\omega)  
\end{equation}
which is  a Fourier cosine  transform of the response function $\mathbf{R}(t)$!  In turn, its inverse Fourier  transform reads 
\begin{equation} \label{PI}
 \mathbf{R}(t) = \int_0^{\infty} d\omega \;  \hat{\mathbf{R}}_F(\omega) \cos(\omega t). 
\end{equation}
From the theory of Laplace transform it follows that $\lim_{z\to\infty} \hat{f}(z) = 0$ for any function $f(t)$ for which the Laplace transform exists. In particular, it is also true for the function $f(t) = d\mathbf{R}(t)/dt$. Calculating its Laplace transform, we obtain  the relation
\begin{equation}
\mathbf{R}(0) = \lim_{z\to\infty} z \hat{\mathbf{R}}_L(z) = 1.
\end{equation} 
On the other hand from (\ref{PI}) we get 
\begin{equation}
\mathbf{R}(0) = \int_0^{\infty} d\omega\; \hat{\mathbf{R}}_F(\omega) = \int_0^{\infty} d\omega\; \mathbb{P}(\omega) = 1.
\end{equation}
 So, we proved that there exists a random variable $\xi$ for which $\mathbb P$ is its probability distribution.  This random variable is interpreted as a  frequency of the thermostat oscillators and Eq.  (\ref{Ek}) is an average value of the function $\mathcal{E}_k(\xi)$ of the random variable $\xi$ (physicists frequently equate it with the integration variable). The probability distribution (\ref{P}) seems to be surprisingly simple in its form as  expressed by the Laplace transform (\ref{RL}) of the response function. The form (\ref{PC}) of  $\mathbb{P}(\omega)$ looks even more simpler: it is a Fourier cosine  transform of the response function $\mathbf{R}(t)$  which solves the GLE in Eq. (\ref{GLE}).

\begin{figure}[t] 
	\centering
	\includegraphics[width=0.45\linewidth]{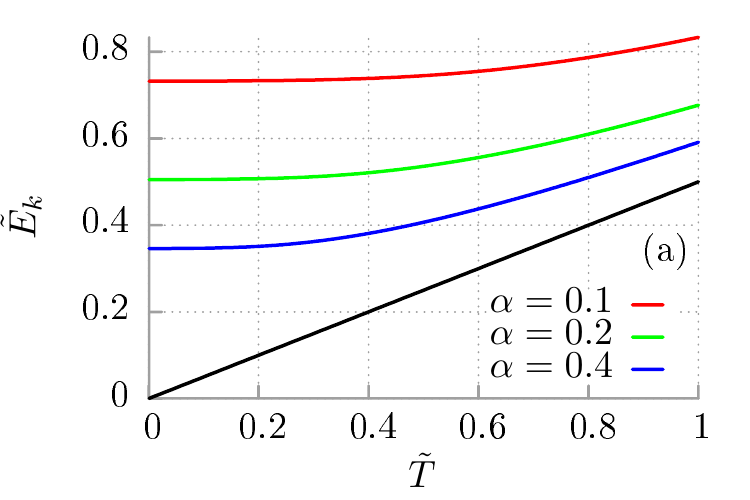}
	\includegraphics[width=0.45\linewidth]{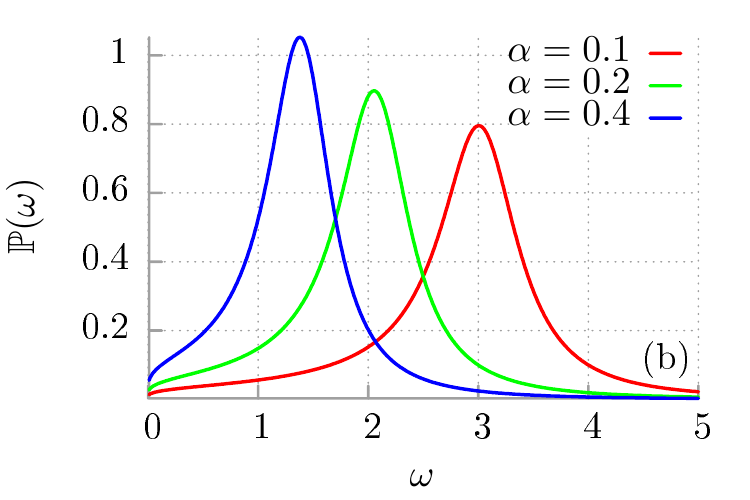}
	\caption{Algebraic decay of the dissipation function \mbox{$\gamma(t) = \gamma_0/(t+\tau_c)$} with the coupling $\gamma_0$ and the memory time $\tau_c$. Panel (a): Average kinetic energy of the free quantum Brownian particle of a mass $M$ as a function of rescaled temperature is depicted for three values of the  dimensionless parameter $\alpha = M/(\gamma_0 \tau_c)$. The rescaled energy is $\tilde E_k=E_k/\hbar \varepsilon$  and the rescaled temperature is $\tilde T=k_B T/\hbar \varepsilon$ with $\varepsilon =1/\tau_c$. The black curve corresponds to the classical case 
	$\tilde E_k = \tilde T/2$. Panel(b): The probability distribution $\mathbb{P}(\omega)$ for selected values of $\alpha=M/\tau_c \gamma_0$.}
	\label{fig1}
\end{figure}

In the following, we consider two examples of the memory kernel $\gamma(t)$, namely the algebraically decaying dissipation function 
\begin{equation} \label{gamma-a}
\gamma(t) = \frac{\gamma_0}{t+\tau_c} 
\end{equation}
and the exponentially decaying oscillations \cite{bialas} 
\begin{equation} \label{gamma-e}
\gamma(t) = \frac{\gamma_0}{\tau_c} e^{- t/\tau_c} \cos (\Omega t),  
\end{equation}
where the parameter $\gamma_0$ is the strength of the system-thermostat coupling and $\tau_c$ defines decay or relaxation time which characterizes memory effects. The case of $\Omega = 0$ in Eq. (\ref{gamma-e}) corresponds to the so-called Drude model of quantum dissipation. In Fig. \ref{fig1} (a) we present the kinetic energy $E_k$ of the free quantum Brownian particle versus thermostat temperature $T$ for the algebraically decaying damping kernel (\ref{gamma-a}).   A more deeper analysis of  $E_k$ is performed in Ref. \cite{bialas} for the case  (\ref{gamma-e}) (although in a different context, see also  Fig. 3b  of  Ref. \cite{ingold} for a harmonic oscillator with the Drude model of dissipation). It supports the very well known result that  kinetic energy of the quantum Brownian particle is always greater than the classical one and at zero temperature $T=0$ the kinetic energy $E_k>0$.

What is indeed crucially new is shown in panel (b) of Fig. 1 and in Fig. 2 for  $\gamma(t)$  given by  Eq. (\ref{gamma-e}).  We can note that the thermostat oscillators of some frequencies, say $\tilde{\omega}$, contribute to $E_k$ with much greater probability then the others. In panel (b) of Fig.1, the most probable frequency $\tilde{\omega}$ depends on the mass $M$ of the Brownian particle, the coupling strength  $\gamma_0$ and  the memory time $\tau_c$. However, it does not depend on these three parameters separately but only on their specific combination  $\alpha=M/(\tau_c \gamma_0)$.  
 For small values of the dimensionless parameter $\alpha$ (i.e. for long memory time or/and strong coupling), 
 mainly high-frequency-oscillators contribute to $E_k$.   In turn, for large values of  $\alpha$ (i.e. for short memory time or/and weak coupling),  mainly low-frequency-oscillators  contribute to $E_k$. 
\begin{figure}[t]
	\centering
	\includegraphics[width=0.45\linewidth]{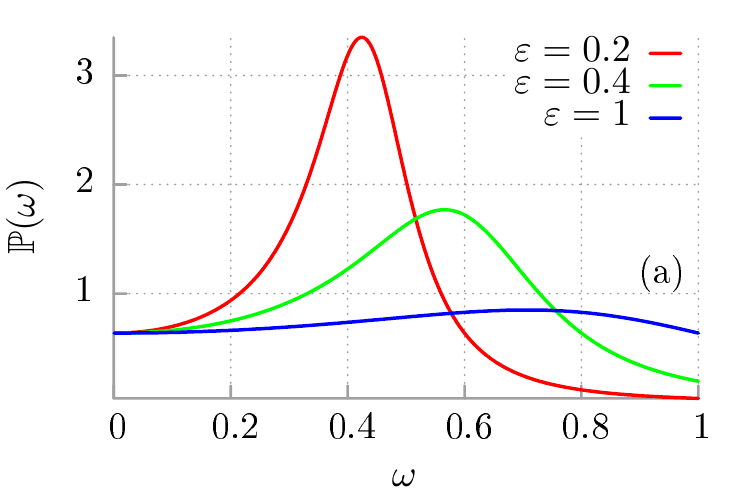}
	\includegraphics[width=0.45\linewidth]{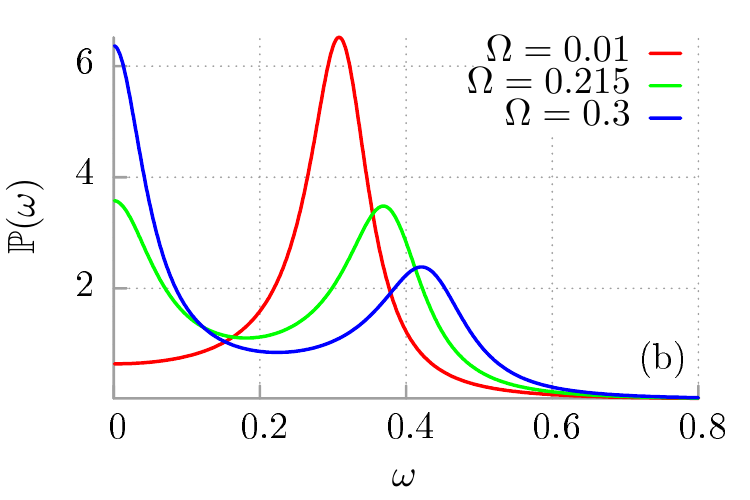}
	\caption{Exponentially decaying oscillation of the dissipation function $\gamma(t) = (\gamma_0/\tau_c)\,e^{- t/\tau_c} \cos (\Omega t)$ with the oscillation frequency $\Omega$. Panel (a): The probability distribution $\mathbb{P}(\omega)$ is presented for Drude model $\Omega = 0$ and different values of the memory time $\varepsilon = 1/\tau_c$. Panel (b): The impact of the oscillation frequency $\Omega$ on the probability density function $\mathbb{P}(\omega)$ with $\varepsilon = 0.1$.}
	\label{fig2}
\end{figure}
A radically different behaviour is visible for the exponentially decaying oscillations in the dissipation function given by Eq. (\ref{gamma-e}). In this case, there are three characteristic frequencies $\varepsilon = 1/\tau_c, \mu_0 = \gamma_0/M$ and $\Omega$. In panel (a) of Fig. \ref{fig2}, the case of the Drude model $\Omega = 0$ is considered and the role of the memory time $\tau_c$ is depicted. We can see that for the long memory time  the probability distribution $\mathbb{P}(\omega)$ is very narrow with high peak at some low frequency $\omega$. When $\tau_c$ decreases ($\varepsilon$ increases) the half-width of $\mathbb{P}(\omega)$ increases. It means that oscillators of a larger frequency interval contribute significantly to the kinetic energy $E_k$. Finally, for small memory time the distribution is almost flat indicating that oscillators of almost all frequencies bring nearly equal contribution to $E_k$. 

The qualitatively new features emerge for the oscillatory dissipation function with $\Omega > 0$. For small $\Omega$ the probability distribution $\mathbb{P}(\omega)$ is unimodal and has one maximum visible for higher frequencies $\omega$. When $\Omega$ increases a new extremum is born at zero frequency $\omega = 0$. This means that modes of low frequency starts to significantly contribute to the kinetic energy $E_k$. For intermediate values of $\Omega$ the probability distribution $\mathbb{P}(\omega)$ exhibits a clear bimodal character. Then both modes of low and high frequency play a crucial but equal role. Further increase of $\Omega$ extinguishes the contribution of higher frequencies at the favour of the near zero frequency modes which are then the most pronounced ones.

Now, let us make a comment about the role of statistical moments $\langle \xi^n \rangle = \int_0^{\infty} d\omega\, \omega^n \mathbb{P}(\omega)$ of the random variable $\xi$ distributed according to the probability density $\mathbb{P}(\omega)$. A caution is needed since not all moments may exist. At least the first two moments have to be finite because they have a clear physical interpretation. The first moment (i.e. the mean value of the distribution) $\langle \xi \rangle$ is proportional to the kinetic energy $E_k$ at zero temperature $T=0$, namely, 
\begin{equation} \label{E0}
E_k(T=0) = \frac{\hbar}{4} \langle \xi \rangle. 
\end{equation}
The second moment $\langle \xi^2 \rangle$ is proportional to the first correction of $E_k$ in the high temperature regime, 
\begin{equation} \label{ET}
E_k = \frac{1}{2} k_B T + \frac{\hbar^2}{24k_BT} \langle \xi^2 \rangle.  
\end{equation}

\section*{Discussion}

In this part  we want to make  the following  statements and remarks: 
\begin{itemize}
\item We have derived the same expressions (\ref{Ek}) and  (\ref{P}) by the same method  for a quantum harmonic oscillator. In this case, one has to replace Eq. (\ref{RL}) by the response function of the harmonic oscillator, namely, 
\begin{equation} \label{oscy} 
\hat{\mathbf{R}}_L(z) = \frac{Mz}{Mz^2 + z\hat \gamma_L(z) + M\omega_0^2}. 
\end{equation}
\item  Eq. (\ref{Ek}) can be generalized for any quantum systems for which: \\
-- the relation 
$p(t) \propto  \int_0^{t} du \; \mathbf{R}(t-u) \eta(u)$ is satisfied,\\
-- the fluctuation-dissipation relation $\hat{C}_F(\omega)= \mathcal{E}(\omega)  \times \hat{\gamma}_F(\omega)$ is satisfied with some function  $\mathcal{E}(\omega)$.  
\item From the above it follows that (\ref{Ek}) holds true within the linear response theory and can be obtained from the fluctuation-dissipation relation of the Callen-Welton type \cite{breuer,weis,call,kub,landau}. We recall that in this theory the quantum system is characterized by the Hamiltonian  $\tilde H$  and is in a thermal equilibrium state at temperature $T$ defined by the Gibbs canonical statistical operator $\rho \propto \mbox{exp}[-\tilde H/k_B T]$. Next, the external force $F(t)$ is applied to the system which develops in time under the perturbed time-dependent Hamiltonian $H=\tilde H- F(t) A$, where $A$ is a hermitian operator. In the linear response approximation  one can calculate fluctuations of the operator $B$. For a special choice of $A$ and $B$ one  can get Eq. (\ref{Ek}).  We quote two examples: 

1.  In Ref. \cite{ford}, the authors have obtained  Eq. (4.14):  
\begin{equation}
	\label{ford}
	E_k = \frac{\hbar}{2\pi} \int_0^\infty d\omega \, \coth{\left[\frac{\hbar\omega}{2k_BT}\right]}\,M\omega^2 \,  \mbox{Im} [\alpha(\omega + i 0^+)].  
\end{equation}
By comparing  this equation  with our formulas we see that $\mathbb{P}(\omega) = (2/\pi) M\omega \, \mbox{Im} [\alpha(\omega + i 0^+)]$, where  $\alpha(\omega)$ is called susceptibility.  

2.  The second example is Eq. (124.10) in the Landau-Lifshitz book \cite{landau} which for  the momentum operator takes the form: 
\begin{equation}
	\label{landau}
	\langle p^2 \rangle = \frac{\hbar}{\pi}\int_0^\infty d\omega\, \coth{\left[\frac{\hbar\omega}{2k_BT}\right]} \,\chi''(\omega),
\end{equation}
where $\chi''(\omega)$ is the imaginary part of the generalized susceptibility $\chi(\omega) = \chi'(\omega) + i\chi''(\omega)$.  
If one knows the relation (\ref{Ek}) together with Eq. (\ref{ho}) and compare Eq. (\ref{landau})  with them then  one  obtains the relation  
\begin{equation} \label{chi}
\mathbb{P}(\omega) = \frac{2}{\pi M \omega } \, \chi''(\omega).
\end{equation}
Also other formulas cited in literature can be re-formulated to the form (\ref{Ek}).

\item Applying this fluctuation-dissipation relation we see that Eq. (\ref{Ek}) is valid for arbitrary systems in contact with bosonic thermostat. It means that Eq. (\ref{Ek}) holds true for any potential $U(x)$ in the Hamiltonian (\ref{H}). For this class of systems the  quantum partition of kinetic energy  (\ref{Ek}) is universal and  the probability distribution is of the form  (\ref{chi}), where the susceptibility $\chi''(\omega)$ is the imaginary part of the Fourier transform of the response function $\chi(t)$. It can be calculated by the method of e.g. the retarded thermodynamic Green functions \cite{zubarev}:  
\begin{equation} 	\label{green} 
	\chi(\omega) = \int_{-\infty}^{\infty} dt\, \mbox{e}^{i\omega t} \,\chi(t), 
			\quad 
		\chi(t) =  \frac{i}{\hbar} \theta(t) \langle [p(t), p(0)] \rangle,
\end{equation}
where $\theta(t)$ is the Heaviside step function,  $p(t) = \exp(i{\tilde H} t/\hbar) p(0) \exp(-i{\tilde H} t/\hbar)$ and  averaging is over the Gibbs canonical statistical operator $\rho \propto \mbox{exp}[-\tilde H/k_B T]$. If $\tilde H = H$ with total $H$ as in (\ref{H}) then all regimes, from weak to strong coupling with thermostat, can be analyzed. However, if $\tilde H = p^2/2M + U(x)$ then only the weak coupling limit can be considered. 

\item The above mentioned formula (\ref{chi}) 
establishes the relation between the probability distribution $\mathbb{P}(\omega)$ and the generalized susceptibility $\chi''(\omega)$. It means that properties of  the quantum environment and its coupling to a given quantum system which are characterized  by $\mathbb{P}(\omega)$ may be experimentally inferred from the measurement of the linear response of the system to an applied perturbation,   for instance electrical or magnetic. Consequently, the latter quantity may open a new pathway to study quantum open systems.
\item Let us observe that  Eq. (\ref{Ek}) can be interpreted in the framework of superstatistics \cite{beck}. Indeed, the quantity $\mathcal E_k(\omega)$ is additionally averaged over the random variable $\xi$, i.e. over randomly distributed frequencies of the thermostat oscillators according to the probability density $\mathbb{P}(\omega)$. 
\end{itemize}

In conclusion, we disclose a  new face  of the old and well known relations of quantum statistical physics. We propose the quantum analogue for partition of kinetic energy in the case of  two exactly solvable systems, namely, a free quantum particle and a quantum harmonic oscillator. There are three new elements: the probabilistic form of Eq. (\ref{Ek}), its interpretation as  mean kinetic energy of the thermostat degree of freedom and information embodied in  $\mathbb{P}(\omega)$. 
The proposed re-interpretation of the relation (\ref{Ek})  has a transparent and intuitive meaning as in the classical case: The mean kinetic energy of a Brownian  particle equals the mean  kinetic energy of thermostat per one degree of freedom.  The relation (\ref{P}) is valid for arbitrary values of the system-thermostat coupling, from weak coupling to strong coupling regimes.  A particularly simple is the form of 
  the probability density function $\mathbb{P}(\omega)$ in the representation (\ref{PC}), which is the Fourier cosine transform of the response function $\mathbf{R}(t)$. It is a challenge to  extend our approach to other quantum systems and to other (spin, fermionic, ...)  environments to show that {\it mutatis mutandis} the counterpart of  (\ref{Ek}) is universal and holds true for all quantum systems. The only non-trivial problem is to determine $\mathbb{P}$. We hope that the present work stimulates such further theoretical analysis.  
\section*{Acknowledgement}
The work  supported by the Grants NCN 2015/19/B/ST2/02856 (P. B. and J. {\L}.) and NCN 2017/26/D/ST2/00543  as well as  the Foundation for Polish Science (FNP) Start fellowship (J. S.).

\end{document}